\documentclass[letterpaper,twocolumn,10pt]{article}
\usepackage{usenix,epsfig,endnotes,hyperref}
\usepackage[square]{natbib}
\usepackage{latexsym}
\usepackage{amsmath}
\usepackage{amssymb}
\usepackage{epsfig}
\usepackage{latexsym}
\usepackage{setspace}
\usepackage{mathrsfs}

\newcommand{\cL}{{\mathcal{L}}}
\newcommand{\cW}{{\mathcal{W}}}

\begin{document}

\date{26 June 2011}

\title{\Large \bf SOBA: Secrecy-preserving Observable Ballot-level Audit}

\author{Josh Benaloh, Microsoft Research\\
            Douglas Jones, Department of Computer Science, University of Iowa\\
            Eric L.~Lazarus,  DecisionSmith\\
            Mark Lindeman\\
            Philip B.~Stark, Department of Statistics, University of California, Berkeley
}

\maketitle

\abstract{
SOBA is an approach to election verification that provides observers with justifiably 
high confidence that the reported results of an election are consistent with an
audit trail (``ballots''), which can be paper or electronic. 
SOBA combines three ideas: 
(1)~publishing cast vote records (CVRs) separately for each contest, 
so that anyone can verify that each reported contest outcome is correct, 
if the CVRs reflect voters' intentions with sufficient accuracy; 
(2)~shrouding a mapping between ballots
and the CVRs for those ballots to prevent the loss of privacy 
that could occur otherwise; 
(3)~assessing the accuracy with which the CVRs reflect voters' intentions for a 
collection of contests while simultaneously assessing the integrity of the 
shrouded mapping between ballots and CVRs by comparing randomly selected 
ballots to the CVRs that purport to represent them.  
Step~(1) is related to work by the Humboldt County Election Transparency Project, 
but publishing CVRs separately 
for individual contests rather than images of entire ballots preserves privacy. 
Step~(2) requires a cryptographic commitment from elections officials.  
Observers participate in step~(3), which relies on the 
``super-simple simultaneous single-ballot risk-limiting audit.'' 
Step~(3) is designed to reveal relatively few ballots if the shrouded 
mapping is proper and the CVRs accurately reflect voter intent. 
But if the reported outcomes of the contests differ from the outcomes 
that a full hand count would show, 
step~(3) is guaranteed to have a large chance of requiring all the ballots to
be counted by hand, 
thereby limiting the risk that an incorrect outcome will become 
official and final.
}

\section{Introduction and background}
The majority of Americans now vote electronically, 
either on machine-counted paper ballots or on 
Direct Recording Electronic (DRE) machines.  
Electronic voting offers advantages over hand counts and lever machines, but it poses challenges 
for determining whether votes were recorded and counted correctly. 
A wide range of security vulnerabilities and other flaws have been 
documented in contemporary voting equipment. 
The 2007 ``Top-to-Bottom Review'' of the systems used in 
California found that all the systems had ``serious design flaws'' 
and ``specific vulnerabilities, which attackers could exploit to 
affect election outcomes''~\citep{bowen07}. 
While some of these vulnerabilities can be mitigated, the underlying verification challenge is formidable. 
As Rivest and Wack comment, ``complexity is the enemy of security,'' and demonstrating that any 
complex system is free of faults may be impossible or infeasible~\citep{rivestWack06}.

Electronic voting systems have failed in real elections. 
In the 2004 general election in Carteret County, North Carolina, over 
4,000 votes were lost irretrievably due to a programming error that affected
UniLect Patriot voting machines, casting doubt on a statewide 
election outcome~\citep{bonner04}. 
More controversially, in the 2006 general election, 
ES\&S iVotronic DREs in Sarasota County, Florida did not record a vote for 
U.S.~House for about 15\% of voters---far more than 
can plausibly be attributed to intentional undervoting.
Inadvertent undervotes were probably decisive in that 
contest~\citep{ashLamperti08,mebaneDill07}. 
Hypotheses explaining these undervotes include voter confusion 
caused by poor ballot layout~\citep{frisinaEtal08} and machine failure~\citep{garber08,mebane09}.  
Unfortunately, the forensic evidence generated by the voting systems
was inadequate to determine the cause of the undervotes or the intentions of the
voters.

Voter-marked paper ballots provide a clearer record of what voters did and more evidence 
about voter intent, but by themselves do not solve the election verification problem. 
In 2005, Harri Hursti repeatedly demonstrated the ability to ``hack'' optical scan counts 
when given access to a memory card~\citep{zetter05}. 
In a June 2006 primary election in Pottawattamie County, Iowa, incorrectly configured optical 
scanners miscounted absentee ballots in every contest, altering two outcomes. 
The county auditor ordered a hand recount, which corrected the errors~\citep{flaherty06}. 
Similar errors in other elections may have altered outcomes without ever being detected. 
Even when scanners work correctly, their results may differ materially from voter intent. 
Consider the 2006 U.S.~Senate contest in Minnesota, where Al Franken beat Norm Coleman in a 
hand recount largely because of ballots where the human interpretation 
differed from the machine interpretation.\footnote{%
   The 2000 presidential election may have been decided by differences 
   between the machine interpretation 
   of certain Florida optical scan ballots and the likely human 
   interpretation~\citep{keating02}.
}

\subsection{Software independence}

Computerized election equipment cannot be infallible, so 
\citet{rivestWack06}
and \citet{rivest08}
suggest that voting systems should be software-independent. 
A voting system is {\em software-independent\/}
``if an undetected change or error in its software cannot cause an undetectable 
change or error in an [apparent] election outcome.''
This idea can be generalized to define independence from hardware and from
elections personnel, leading to so-called {\em end-to-end verifiable\/} 
election technologies.  
However, end-to-end technology may require fundamental 
changes in current voting processes.

The outcome of a contest is the set of winners, not the exact vote counts.
The {\em apparent\/} outcome of a contest is the winner or 
winners according to the voting system.
The {\em correct\/} outcome of a contest is the winner or winners
that a full hand count of the ``audit trail'' would find.
The audit trail is assumed to be an indelible record of how voters cast
their votes.
It might consist of a combination of voter-marked paper ballots, 
voter receipts, a voter-verifiable paper audit trail (VVPAT),
and suitable electronic records.

This definition of ``correct'' is generally a matter of law. 
It does not necessarily imply that the audit trail is inviolate
(nor that the outcome according to the audit trail is the 
same as the outcome according to how voters originally cast their ballots);
that there is no controversy about which records in the audit trail reflect
valid votes;
that human observers agree on the interpretation of the audit trail; 
that the actual hand counting is accurate; 
nor that repeating the hand count would give the same answer.
If there is no audit trail, defining what it means for the apparent 
outcome to be correct requires hypothetical counterfactuals---but for 
the fault in the voting system, what would the outcome have been?

Software independence means that errors that cause apparent 
outcomes to be wrong leave traces in the audit trail.
But software independence does not guarantee any of the following:
\begin{enumerate}
   \item that no such traces will 
            occur if the apparent outcome is correct\footnote{%
     False alarms are possible.
     An analogy is that if a tamper-evident seal shows that a package has been opened, it does not
     follow that the package contents have been altered.
}
    \item that those traces will be noticed or acted upon
    \item that the cost of looking through the audit trail for those traces is affordable
    \item that, in principle, there is a way to correct the 
             apparent outcome without holding another election
    \item that, in practice, the audit trail was preserved and protected well enough to
             determine the outcome according to how the voters originally 
             cast their ballots
\end{enumerate}
The penultimate property is guaranteed by strong software independence.
\citet{rivestWack06} and \citet{rivest08} define a voting system to be 
{\em strongly software-independent\/}
if an undetected change or error in its software cannot cause an undetectable 
change or error in an [apparent] election outcome, and moreover, a detected change or 
error in an [apparent] election outcome (due to change or error in the software) can be 
corrected without re-running the election.
Having an audit trail does not guarantee that anyone will dig through it to see whether there
is a problem or to correct the outcome if the outcome is wrong. 
Strong software independence does not correct anything, but it is an essential ingredient for
a system to be self-correcting.

{\em Compliance audits\/} can be used to assess whether the last property listed above holds: 
Given that the election used a strongly software-independent voting system,
did it adhere to procedures that should keep the audit trail sufficiently 
accurate to reconstruct the outcome according to how voters cast their ballots?
Strong evidence that such procedures were followed is strong evidence that
the legally correct outcome---what a full hand count of the audit trail would show---is 
the same as the outcome according to how the voters originally cast their ballots.
As we discuss below in section~\ref{sec:discussion},
we believe that compliance audits should always be required:
If the election fails the compliance audit,\footnote{%
    ``Failure'' means failure to find strong evidence that such procedures were followed,
    rather than finding evidence that such procedures were not followed.
}
there is no assurance that even a full hand
count of the audit trail would show the outcome according to
how the voters really voted.
Below, we assume that the election has passed a compliance audit.

\subsection{Vote tabulation audits}
Vote tabulation audits compare reported vote subtotals for subsets of ballots
(``audit units'')
with hand counts of the votes for each of those subsets.
Audit units have to be subsets for which the voting system reports vote subtotals.
Most present U.S.~audits use audit units that consist of all the ballots cast in
individual precincts or all the ballots tabulated on individual voting machines.
Generally, audit laws do not have provisions that would lead to correcting incorrect 
electoral outcomes~\citep{hallEtal09}.\footnote{%
    For instance, under New York law, each county determines independently 
    whether its audit in a particular contest must be expanded. 
    This provision means that a correct outcome might be changed to an 
    incorrect outcome even if the conduct of the audit is formally flawless.
}

A {\em risk-limiting post-election audit\/} uses the audit trail 
to guarantee that there is a large, pre-specified probability that the 
audit will correct the apparent outcome if the apparent outcome is wrong.
Risk-limiting audits are widely considered best practice~\citep{bestPractices08}.
Risk-limiting audits have been endorsed by 
the American Statistical Association~\citep{asa10}, 
the Brennan Center for Justice, 
Common Cause, 
the League of Women Voters, 
and Verified Voting, among others.
California AB~2023 (2010), requires a pilot of risk-limiting audits in 
2011~\citep{ab2023_2010}.
Colorado Revised Statutes \S1-7-515
calls for implementing risk-limiting audits by 2014.

The first method for conducting risk-limiting audits was 
proposed by~\citet{stark08a}; 
numerous improvements have been 
made~\citep{stark08d,stark09a,stark09b,stark09d,miratrixStark09a,stark10d}.
See also~\citep{checkowayEtal10}.
Risk-limiting audits limit the risk of failing to correct an outcome that is wrong.
The risk limit is 100\% minus the minimum chance that the 
audit corrects the outcome.
If the outcome is correct in the first place, a risk-limiting audit 
cannot make it wrong;
but if the outcome is wrong, a risk-limiting audit has a large chance of correcting it.
Hence, the probability that the outcome according to a 
risk-limiting audit is the correct outcome
is at least 100\% minus the risk limit.

For systems that are strongly software-independent, 
adding a risk-limiting audit addresses the second condition above: 
It ensures a large, pre-specified probability that the traces 
will be noticed and will be used to correct the
apparent outcome if the apparent outcome is wrong.

\subsection{Our goal}
Our goal in this work is to sketch a personally verifiable
privacy-preserving $P$-resilient canvass framework.  
We must first say what this means.

A {\em canvass framework\/} 
consists of the vote-tabulation system together with
other human, hardware, software, and
procedural components of the canvass, including compliance and vote-tabulation audits.
A canvass framework is {\em resilient with probability $P$\/} or 
{\em $P$-resilient\/} if the 
probability that the outcome it gives\footnote{%
   As discussed in section~\ref{sec:discussion},
   to be $P$-resilient, a canvass framework should refrain from giving any outcome at all
   if some preconditions are not met.
} 
is the correct outcome is at least $P$, 
even if its software has an error,
shortcoming, or undetected change.\footnote{%
    The probability comes from the overall voting system, in our case from the fact that
    the audit relies on a random sample.
    The probability does not come from treating votes, voters, or election outcomes 
    as random, for instance.
}
Resilience means that the framework tends to recover from faults.
If a canvass framework is $P$-resilient, either the
outcome it gives when all is said and done is correct, 
or something occurred that had probability less than $1-P$.
The canvass framework that results from performing a risk-limiting audit
on a strongly software-independent voting system that passes a compliance audit
is $P$-resilient, with $P$ equal to 100\% minus the risk limit.
If the system fails the compliance audit, the framework should not declare any outcome.
Instead, the election should be re-run.

Even if a canvass framework is $P$-resilient, in practice the public might not 
trust the system unless they can observe crucial steps, 
especially the audit.
The mere right or opportunity to observe the audit will not engender much 
trust if---as a practical matter---no single person or small group {\em could\/} observe all
the steps that are essential to ensuring the accuracy of the final result.
For instance, if a vote-tabulation audit takes ten teams of auditors working in separate
offices four days to complete, it would take a large team of independent observers---with
lots of free time and long attention spans---to verify that the 
audit was carried out correctly.
The longer an audit takes and the more people required to carry out the audit, 
the more opportunities there are to damage the audit trail, and the 
harder it is for an observer to be satisfied that the audit has been conducted correctly. 

We define a canvass framework to be {\em personally verifiable $P$-resilient\/} 
if it is $P$-resilient and a single individual could, as a practical matter, 
observe enough of the process to have convincing evidence that the
canvass framework is in fact $P$-resilient.

The transparency required for a canvass framework to be personally 
verifiable can impact privacy.
For instance, publishing images of all the ballots cast in an election\footnote{%
   There also needs to be proof that the images are sufficiently complete and accurate
   to determine the correct outcome.
}
might give the individuals compelling evidence that the vote tabulation system found the 
correct outcome, since the images allow people to count the 
votes themselves---at least to the extent that voter intent is unambiguous.\footnote{%
       Verification methods like Humboldt County Election Transparency Project (see below) 
       involve publishing digital images of all the ballots. 
}
But publishing ballot images can facilitate vote-selling and coercion and 
can compromise privacy, because voters can deliberately or 
accidentally reveal their identities through marks on the ballots including
idiosyncrasies of how individuals fill in 
bubbles~\citep{calandrinoEtal11} or even 
the fiber structure of the paper on which the ballot is 
printed~\citep{calandrinoEtal09}.\footnote{%
    There are arguments that images of ballots should be published anyway---that
    transparency is more important than privacy.
    In jurisdictions that permit voting by mail, there is an opportunity to confirm
    how someone votes for the purpose of vote-selling or coercion; indeed, someone
    could fill out another's ballot.
    Whether publishing images of ballots would change the rate of vote-selling
    or coercion substantially is the subject of some debate.
}

A lesser but substantial degree of transparency is conferred by 
publishing cast vote records (CVRs)\footnote{%
   In the 2002 FEC Voting System Standards~\citep{FEC2002b}, 
   these were called ``ballot images'';
   however, the term CVR has been used in more recent EAC Voluntary Voting System
   Guidelines~\citep{EAC2005b}; we prefer the latter 
   term because it does not suggest an actual image
   but rather a record of the interpretation of the system's interpretation of the ballot.
   And what matters is the system's interpretation of the ballot as a set of votes.
} 
enabling anyone to verify that the contest outcomes are 
correct---if the CVRs are accurate.
However, as~\citet{popoveniucStanton07} and \citet{rescorla09} point out, 
publishing CVRs also can aid vote-selling or coercion because of the 
potential for pattern voting. 
One typical sample ballot (from Tulsa, Oklahoma) contains 18~contests 
with over 589,000 possible 
combinations if a voter votes in every contest, or over 688~million 
combinations allowing for undervotes. 
Thus, a voter could be instructed to vote for the preferred 
candidate in one contest, and to cast a 
series of other votes that would almost certainly (especially within a precinct), 
confirm the voter's identity if all of the voter's selections were published.
Hence, publishing whole-ballot CVRs for large numbers of 
ballots improves transparency but can sacrifice privacy.

When there is not strong evidence that the apparent outcome is correct,
risk-limiting audits can require examining the entire audit trail, 
potentially exposing all the ballots to public scrutiny.\footnote{%
   One could have a risk-limiting audit that, if it had not terminated
   after some fraction of the ballots had been examined, triggered a hand count
   of the remaining ballots, but did not allow the public to observe that hand
   count.
   But then why should the public trust that the hand count was accurate?
}
If the apparent outcome is wrong, such exposure is necessary 
in order to correct the outcome.
Therefore, if a risk-limiting audit is to be personally verifiable, 
there may be occasions where 
compromising privacy is unavoidable.
But minimizing the number of ballots or whole-ballot CVRs that are 
routinely exposed helps protect privacy, impeding vote-selling and coercion.

We define a canvass framework to be 
{\em personally verifiable privacy-preserving $P$-resilient\/}
if it is personally verifiable $P$-resilient and it does not sacrifice privacy unnecessarily.
Neither {\em personally verifiable\/} nor {\em privacy-preserving\/} is 
a mathematically precise characteristic, while $P$-resilience is.

The contribution of the present work is to sketch a personally verifiable privacy-preserving
$P$-resilient voting system.
We assume, as a foundation for building this system,
that we are starting with a strongly 
software-independent voting system with an audit trail that corresponds 
to individual ballots.
Moreover, we assume that a compliance audit has determined that the 
audit trail generated by the system is sufficiently trustworthy to reflect the correct outcomes of
the contests.
We augment the system with procedures and data structures that make it possible for
an individual observer to gain compelling evidence that either the outcomes are correct, or
something very unlikely occurred---that is, that the overall canvass framework is $P$-resilient.
Unless some of the apparent outcomes are wrong or a margin is extremely small, gathering
that evidence will generally involve exposing only a tiny percentage of ballots and 
whole-ballot CVRs.

In essence, our method adds a special risk-limiting audit to a strongly 
software-independent voting system (one that has had a compliance audit to ensure
that its audit trail is intact).
Since one person cannot be in two places at the same time, the procedure 
cannot be personally verifiable if it involves auditing a multi-jurisdictional contest 
in different jurisdictions simultaneously; it would then be necessary to trust 
confederates to observe what is happening elsewhere.
The next few sections outline elements of this risk-limiting audit.

\section{Ballot-level risk-limiting audits}
One key to keeping the process personally verifiable 
(by keeping amount of observation required low)
and to protecting privacy 
(by exposing as few ballots as possible to observers) is
to audit the record at the level of individual ballots, rather than large batches of
ballots such as precincts.
The fewer ballots there are in each audit unit, the smaller the expected counting 
burden for risk-limiting audits tends to be---when the electoral outcome 
is correct (see, e.g., \citep{stark09c,stark10c,stark10d}).
A vote-tabulation audit based on checking the CVRs of individual 
ballots against a human interpretation of those ballots
is often called a ``ballot-level audit,''
a ``single-ballot audit,'' or a ``ballot-based audit.'' 
Because they reduce the time it takes to audit and the number of ballots
involved, ballot-level risk-limiting audits are especially amenable to 
personal verification.

Ballot-level audits are extremely efficient statistically, but they are not
simple to implement using current voting systems.
To perform a ballot-level audit, there must be a way to identify each 
ballot uniquely, for instance, a serial number on a paper ballot, 
or identifying the ballot by its location:
``the 17th ballot in deck 152 scanned by scanner~C,'' for instance.\footnote{%
  If an identifier is printed on paper ballots, the printing 
  should occur after the voter casts his or her vote
  and the ballots are co-mingled.
  If the identifier is printed before the voter casts his or her vote, privacy 
  could be compromised.
}
There must also be a way to match each ballot to its CVR.
Some commercial voting systems do not generate or do not store 
CVRs for individual ballots.
Other voting systems record individual CVRs, but are
designed make it difficult or impossible to match individual CVRs to the ballots they
purport to represent. 
In some cases, audit trails have identifiers that can be used to find the corresponding CVRs; 
this method was used for part of a 2008 audit in Eagle County, 
Colorado~\citep{branscomb08}\footnote{%
   Optical-scan ballots as well as DRE paper audit trails can have identifiers. 
   For instance, in Boulder County, Colorado, the Hart Ballot Now system is 
   configured to print unique identifiers and bar codes on each ballot.
   In Orange County, California, ballots for the Hart Ballot Now system have non-unique 
   identifiers and bar codes (numbered 1--2500, then repeating).
}
and a ballot-level risk-limiting audit in Orange County, California, in 2011 
[P.B.~Stark, personal communication, 2011]. 
However, to protect privacy, most paper ballots do not have identification numbers. 
In a 2009 pilot ballot-level audit in Yolo County, California, \citet{stark09d} 
exploited the fact that the CVRs and the physical ballots were in the same order. 
The scanned images associated with each CVR in the audit sample were compared with the 
physical ballots to check the accuracy of the CVRs.  

\citet{calandrinoEtal07} 
describe an approach to election verification that  
involves imprinting ballots with identification numbers and scanning the ballots with a 
``parallel'' system in addition to the system of record.
The parallel system derives its own CVRs, from which the apparent
contest outcome can be determined independently.
The accuracy of the unofficial CVRs and of the imprinting process is 
then assessed by a ballot-level audit.

Since 2008, the Humboldt County Election
Transparency Project (Humboldt County ETP) has experimented with publishing 
ballot images and independently tabulating CVRs extracted from those images. 
Using commercially available equipment, Humboldt County ETP rescans paper 
ballots after embossing them with serial numbers. 
Then, open-source software is used to form CVRs from the digital images. 
Humboldt County ETP has processed ballots for six elections and
published scanned ballot images as well as its version of the 
CVRs for some of them.
The results based on their re-scans generally have agreed well with the 
original results, with one important exception: 
The Humboldt County ETP analysis of the November 2008 election uncovered a defect in the 
election management software that led the results of an entire 
ballot batch to be silently discarded!

The Clear Ballot Group, inspired in part by Humboldt County ETP, is 
developing a system that, in its words, could permit election outcomes to be 
``thoroughly and transparently verified within 36--48 hours after the polls close.''
Neither the Humboldt County ETP nor 
Clear Ballot Group currently incorporate risk-limiting audits,\footnote{%
   Clear Ballot Group is adding support for risk-limiting audits to their software
   [L.~Moore, personal communication, 2011].
}
but the parallel scans their systems perform facilitate ballot-level risk-limiting audits, 
along the general lines proposed
by~\citet{calandrinoEtal07}.
If the system of record and the parallel system agree on the set of 
winners, a risk-limiting audit of the parallel system transitively confirms 
the outcome according to the system of record.\footnote{%
   This is true as long as the systems agree on the set of winners, even if they 
   disagree about vote totals or margins.
   For instance, suppose candidate A defeats candidate B by one 
   percentage point in the original returns, 
   and by ten points according to the parallel system.
   Such a large discrepancy might justify close scrutiny, but a 
   risk-limiting audit of the results of the parallel system
   would still provide strong evidence that A defeated B, or would lead to a full
   hand count to set the record straight.
}

\section{A privacy-preserving audit}
The method we propose here presupposes that CVRs are available, either 
from the system of record or from a parallel system.
It publishes all the data contained in the CVRs in a form that 
(1)~still permits all observers to check the contest outcomes on the 
assumption that the CVRs are accurate,
(2)~does not compromise privacy, and
(3)~enables the CVRs to be checked against the audit trail while minimizing the loss of privacy.

In SOBA, election officials make a cryptographic commitment\footnote{%
   See \url{http://en.wikipedia.org/wiki/Commitment_scheme}.
   Cryptographic commitments have two important properties, the binding property and
   the hiding property, discussed in section~\ref{sec:shroud}.
}
to the full set of CVRs
by publishing the CVRs separately for each contest, disaggregating the ballots
(we call these contest-CVRs or CCVRs in contrast to whole-ballot CVRs), 
and a shrouded link between each
CCVR and the ballot it purports to represent. 
Splitting the CVRs into CCVRs and obfuscating the identity of the ballot from
which each CCVR comes eliminates some of 
the information required to identify a voter's ballot style
or to use pattern voting to signal the voter's identity.\footnote{%
   Of course, if there is a contest in which few voters are eligible to vote, 
   eligibility itself is a signal.
}
This makes the procedure privacy-preserving.
But it retains enough information for any observer to check that the apparent outcome agrees
with the outcome according to the CCVRs, for each contest.
That is, there is a known algorithm (the winner algorithm\footnote{%
   For first-past-the-post contests, the winner algorithm just finds who has the most votes.
   Other voting schemes, such as instant-runoff voting (IRV) or ranked choice voting 
   (RCV), have more complicated winner algorithms.
}) that observers can apply to the published CCVRs to calculate the
correct outcome of every contest---provided the CCVRs reflect the ballots 
(more generally, audit trail) accurately enough.
This is part of making the procedure personally verifiable.
Loosely speaking, the required level of accuracy depends on the number of CVRs
that must have errors for the apparent outcome to be wrong:\footnote{%
   In plurality voting, this is the margin or the set of margins between each (winner, loser) pair.
   Defining the margins for IRV and calculating them for a given set of 
   reported results is not simple.
   See~\cite{cary11,magrinoEtal11}.
} 
The fewer ballots that need to be changed to affect the outcome, 
the larger the sample generally will need to be
to attain a given level of confidence that the apparent outcome is correct.

The CCVRs might fail to be sufficiently accurate because
\begin{itemize}
  \item  At least one CCVR and the ballot it purports to represent do not 
         match because human and machine interpretations of voter intent differ 
         (for instance, because the voter marked the ballot
         improperly).  
         This is a failure of the generation of CCVRs.

  \item  At least one CCVR does not in fact correspond to any ballot.  It is an ``orphan.''
           This is a failure of the mapping between ballots and CCVRs.

  \item  More than one CCVR for the same contest is mapped to the 
            same ballot.  It is a ``multiple.''
            This is also a failure of the mapping between ballots and CCVRs.
         
  \item There is no CCVR corresponding to some voting opportunity on a ballot.

\end{itemize}
A failure of the mapping might be the more distressing source of error, since it is a failure on
the part of the election official, but
we must ensure (statistically) that---together---all sources of error did not combine 
to cause the outcome to be wrong.
SOBA uses a risk-limiting audit to assess statistically whether
the winners according to the full audit trail
differs from the winners according to the CCVRs, for all contests under
audit, taking into account all sources of error.
If the outcome according to the CCVRs is incorrect, the audit is very likely to 
proceed to a full hand count of the audit trail, thereby revealing the 
correct outcome.
This provides $P$-resilience.

To make the risk-limiting audit possible, elections officials are required to publish
another file, the {\em ballot style file\/}, which
contains ballot identifiers and lists the contests each of those 
ballots contains.
It does not contain the voters' selections.

The risk-limiting technique we propose is the 
{\em super-simple simultaneous single-ballot risk-limiting audit\/}~\citep{stark10d}.
It is not the most efficient ballot-level audit, but the calculations it requires
can be done by hand, increasing transparency.
It involves drawing ballots at random with equal probability; some more efficient audits
require using different probabilities for different ballots, which is harder to
implement and to explain to the public.
Moreover, this technique allows a collection of contests to be audited simultaneously
using the same sample of ballots.
That can reduce the number of randomly selected ballots that must
be located, interpreted, and compared with CVRs,
decreasing the cost and time required for the audit and thereby increasing transparency.

The following subsections give more technical detail.

\subsection{Data framework and assumptions}
We assume that the audit trail consists of one record per ballot cast.
There are $C$ contests we wish to assess.
The contests might be simple measures, measures requiring a super-majority,
multi-candidate contests, or contests of the form 
``vote for up to $W$ candidates.''\footnote{%
    We do not specifically consider instant-runoff voting or ranked-choice voting here.
    Risk-limiting methods can be extended to such voting methods, 
    but the details are complex.
}
We refer to records in the audit trail as ``ballots.''
A ballot may be an actual voter-marked paper ballot, a 
voter-verifiable paper audit trail (VVPAT),
or a suitable electronic record.

There are $N$ ballots in the audit trail that each contain one or more of the $C$ contests.
Each ballot can be thought of as a list of pairs, one pair for each contest
on that ballot.
Each pair identifies a contest and the voter's selection(s) in that contest, which might
be an undervote or a vote for one or more candidates or positions.
Examining a ballot by hand reveals all the voter's selections on that ballot;
we assume that there is no ambiguity in interpreting each voter's intentions from
the audit trail.

Before the audit starts, the voting system must report results for 
each of the $C$ contests.
The report for contest $c$ gives $N_c$, the total number of 
ballots cast in contest $c$
(including undervotes and spoiled ballots), as well as
the number of valid votes for each position or candidate in 
contest $c$.
Let $M \equiv N_1 + N_2 + \cdots + N_C$ denote the total number 
of voting opportunities on the $N$ ballots.
We assume that the compliance audit assures us (e.g., through ballot accounting)
that the reported values of $N_c$ are accurate, and that the audit trail is trustworthy.
In the present work, we do not consider attacks on the audit trail.

There is a published ``ballot style file.''
Each line in the ballot style file lists a ballot identifier and a list of contests that ballot
is supposed to contain. 
The ballot identifier uniquely identifies a ballot in the audit trail.
The identifier could be a number that is printed on a paper ballot or unambiguous
instructions for locating the ballot (e.g., the 275th ballot in the 39th deck).
There should be $N$ lines in the file, and the $N$ ballot identifiers should be unique.
Because the ballot style file is published, individual can check this for themselves.
Moreover, individuals can check whether the number of lines 
in the ballot style file that list contest $c$ equals
$N_c$, the total number of ballots the 
system reports were cast in contest $c$.

Before the audit starts, the voting system or a parallel system 
has produced a CVR for each ballot.
These are not published as whole-ballot CVRs.
Rather, the CVRs are split by contest to make contest-specific CVRs (CCVRs) that
contain voters' selections in only one contest.
Each whole-ballot CVR is (supposed to be)
split into as many CCVRs as there are contests on the ballot.

The CCVRs for the contests are published in $C$ files, one 
for each contest.
The CCVR file for contest $c$ should contain $N_c$ lines; 
because this file is published, individuals can check this for themselves.
Each line in the CCVR file for contest $c$ lists a voter's selection and a 
shrouded version of the identifier of the ballot that the selection is 
supposed to represent.
The order of the lines in each of the $C$ CCVR files should by shuffled 
(preferably using random permutations) so that whole CVRs cannot be 
reassembled without knowing secret information.\footnote{%
    For example, each CCVR file could be sorted in order of the shrouded ballot
   identifier.
}

The public can confirm whether the contest outcomes 
according to the CCVR files match the voting system's reported outcomes.
If they do not match, there should be a full hand count of any contests with
discrepant outcomes.
We assume henceforth that the outcomes do match, 
but we do not assume the exact vote totals
according to the CCVR files match the reported vote totals.

The data include one more file that is not published, the {\em lookup file\/}.
The lookup file contains $M$ lines, one for each voting opportunity on each ballot.
Each line has three entries: a shrouded ballot identifier, the corresponding unshrouded
ballot identifier, and a number (``salt'') that is used in computing the shrouded
identifier from the unshrouded identifier using a cryptographic commitment 
function, as described below.
(For a review of uses for cryptography in voting, see~\citet{adida06}.)

The salt on the $j$th line of the file is denoted $u_j$.
Each line corresponds to a (ballot, contest) pair:
We can think of $u_j$ as being $u_{ic}$, the salt used to shroud the identity
of ballot $b_i$ in the CCVR file for contest $c$.
The election official will use this file to convince observers that 
every selection on every ballot
corresponds to exactly one entry in a CCVR file, and vice versa.

\subsection{Shrouding} \label{sec:shroud}
The method of shrouding ballot identifiers is crucial to
the approach.
SOBA requires election officials to cryptographically
commit to the value of the ballot identifier that
goes with each CCVR. 
A cryptographic commitment ensures
that the ballot identifier is secret but indelible: 
The
election official can, in effect, prove to observers that a
shrouded identifier corresponds to a unique unshrouded
identifier, but nobody can figure out which unshrouded
identifier corresponds to a given shrouded identifier without
secret information.

The next few paragraphs describe a suggested instantiation of the 
cryptographic commitment.
We assume that ballot identifiers all have the same length.  
If necessary, this can be achieved by padding identifiers with leading zeros.
The commitment function $H()$ must be disclosed publicly and 
fixed for the duration of the election.

Each commitment represents a claim about a voter's selection(s) 
on a given ballot in a given contest.
For each set of selections that any voter made in each contest, 
including undervotes and votes for more than one candidate,
the election official will create a set of commitments.
Each commitment designates the ballot identifier of a ballot that 
the election official claims contains that set of selections in that contest.
To commit to the ballot identifier $b$, the election official
selects a secret ``salt'' value $u$\footnote{%
   To protect voter privacy, it must be infeasible to guess the salts: 
   Each salt should contain many random or pseudo-random bits.
   For the commitment to be effective, the length of all salt values should be fixed and equal.
   See section~\ref{sec:discussion}.
}
and computes the commitment value $y=H(b, u)$.
At a later stage, the official can open the commitment by revealing
$u$ and $b$: Then anyone can verify that the value $y$
revealed earlier is indeed equal to $H(b, u)$.

Loosely speaking, a commitment function must have
two properties, the {\em binding property\/} and the {\em hiding property\/}.
The binding property makes it infeasible for the official
to find any pair $(b', u') \ne (b, u)$ for 
which $H(b', u')=H(b, u)$. 
This provides integrity by helping to ensure
that election officials cannot contrive to have more than
one CCVR for a given contest claim to come from the
same ballot.\footnote{%
  See step~7 of the proof in section~\ref{sec:proof}.
}
The binding property is crucial for $P$-resilience;
indeed, the proof of $P$-resilience requires only that the commitment
have the binding property and that $\{N_c\}_{c=1}^C$ are known.

The hiding property makes it infeasible for anyone
with access only to the shrouded values $H(b, u)$ to learn
anything about which ballot is involved in each commitment.
This provides privacy by helping
to ensure that observers cannot reassemble whole-ballot
CVRs from the CCVR files without extra information.
If observers could reassemble whole-ballot CVRs, that 
would open a 
channel of communication
(pattern voting) for coercion or vote selling.
Ballot identifier $b$ may appear
in multiple commitments since a separate commitment is generated
for each candidate selection on each ballot.
The hiding property ensures
that those collections of commitments do not together reveal
the value of any $b$.
This is crucial for the method to be privacy-preserving.

An HMAC (as described in Federal Information
Processing Standard Publication~198) with a secure hash
function such as SHA-256 (described in Federal Information
Processing Standard Publication~180-2) can be used to instantiate
the commitment function. 
However, since each of the parameters
of the commitment function is of fixed length it is more efficient to
simply use a cryptographic hash function such as SHA-256 directly.
The length of the ballot identifiers does not matter, as long as all
ballot identifiers in the election have the same length.
We recommend that all salt values have equal length, of at least 128~bits.
Our results do not depend on the particular commitment function
chosen, as long as it has both the binding and hiding
properties.\footnote{%
    \citet{menezesEtal96} offers a thorough treatment of hash functions and their
    use for commitments in applications such as digital signatures.
}

We now describe how to perform a risk-limiting
audit that simultaneously checks the accuracy of
the CCVRs, whether each CCVR entry comes from exactly
one ballot, and whether every voting opportunity on every
ballot is reflected in the correct CCVR file.

\subsection{The audit}
The first three steps check the consistency of the CCVRs with the
reported results and the uniqueness of the shrouded identifiers.

\begin{enumerate}
   \item[1.] Verify that, for each contest $c$, there are $N_c$ entries in 
                 the CCVR file for contest $c$.
   
   \item[2.] Verify that, for each contest $c$, the CCVR file shows
                 the same outcome as the reported outcome.

   \item[3.] Verify that the $M = N_1 + \cdots + N_C$
               shrouded ballot identifiers 
                in all $C$ CCVR files are unique.

\end{enumerate}
If step~2 shows a different outcome for one or more contests, those contests (at least) should be
completely hand counted.

Steps~4 and 5 check the logical consistency of the ballot style file with the reported results.
\begin{enumerate}
     \item[4.] Verify that, for each contest $c$, there are $N_c$ 
                    entries in the ballot style file that list the contest.

     \item[5.] Verify that the ballot identifiers in the ballot style file are unique.

\end{enumerate}
If steps~1, 3, 4, or 5 fail, there has been an error or misrepresentation.
The election official needs to correct all such problems before the audit can start.

The remaining steps comprise the statistical portion of the risk-limiting audit, 
which checks whether the CCVRs and the mapping from ballots to CCVRs is 
accurate enough to determine the correct winner.
\begin{enumerate}
      \item[6.] Set the audit parameters:
                 \begin{enumerate}
                     \item Choose the risk limit $\alpha$. 
                     \item Choose the maximum number of samples $D$ to draw;
                              if there is not strong evidence that the outcomes are 
                              correct after $D$ draws, the entire audit trail will be counted by hand.
                    \item Choose the ``error bound inflator'' $\gamma > 1$ and the error tolerance 
                             $\lambda \in (0, 1)$ for the 
                             super-simple simultaneous method~\citep{stark10d} 
                             ($\gamma = 1.01$ and $\lambda = 0.2$ are reasonable values).
                    \item Calculate 
                           \begin{equation}
                                      \rho = \frac{-\log \alpha}{\frac{1}{2\gamma} + 
                                              \lambda  \log(1 - \frac{1}{2\gamma})}.
                            \end{equation}          
                     \item For each of the $C$ contests, calculate the margin of victory $m_c$ 
                              in votes from the CCVRs for contest $c$.\footnote{%
                                      This would be replaced by a different calculation for IRV or RCV contests.
                                      See, e.g., \citet{magrinoEtal11,cary11}.
                                  }
                     \item Calculate the {\em diluted margin\/} $\mu$: the
                              smallest value of $m_c/N$ among the $C$ contests.\footnote{%
                                  The diluted margin controls the sample size.  
                                  If contest $c$ has the smallest value of $m_c/N$ and $N_c$ 
                                  is rather smaller than $N$, it can be more efficient to audit 
                                  contest $c$ separately
                                  rather than auditing all $C$ contests simultaneously.
                           }
                     \item Calculate the initial sample size $n_0 = \lceil \rho/\mu \rceil$.
                     \item Select a seed $s$ for a pseudo-random number generator (PRNG).\footnote{%
                                      The code for the PRNG algorithm should be published 
                                      so that it can be checked and so that, given the seed $s$, 
                                      observers can reproduce the sequence of pseudo-random 
                                      numbers. 
                                      The PRNG should produce numbers that are statistically
                                      indistinguishable
                                      from independent random numbers
                                      uniformly distributed between 0 and 1  (i.e., have large $p$-values)
                                      for sample sizes 
                                      up to millions for a reasonable battery of tests of
                                      randomness, such as the Diehard tests.
                                  }
                              Observers and election officials could contribute
                              input values to $s$ or $s$ could be generated by an observable, 
                              mechanical source of randomness such as rolls of a 10-sided die.
                              The seed should be selected only once.
                 \end{enumerate}
     
     \item[7.]  Draw the initial sample by finding $n_0$ pseudo-random numbers between $1$ and $N$
                     and audit the corresponding ballots:
                    \begin{enumerate}
                          \item Use the PRNG and the seed $s$ to generate $n_0$ pseudo-random numbers,
                                    $r_1, r_2, \ldots, r_{n_0}$.
                          \item Let $\ell_j \equiv \lceil N r_j \rceil$, $j = 1, \ldots, n_0$. 
                                   This list might contain repeated values.
                                   If so, the tests below only need to be performed once for each value,
                                   but the results count as many times as the value occurs in the list.\footnote{%
                                      The auditing method relies on sampling with replacement to limit the risk.
                                   }
                          \item Find rows $\ell_1, \ldots, \ell_{n_0}$ in the ballot style file.
                          \item Retrieve the ballots $b_{\ell_j}$ 
                                   in the audit trail identified by those rows in the ballot style file.
                                   If there is no ballot with identifier $b_{\ell_j}$, pretend in 
                                   step~7(g) below that the ballot showed a vote for the runner-up 
                                   in every contest listed in that row of the ballot style file.
                          \item Determine whether each ballot shows the same contests as its
                                   corresponding entry in the ballot style file.
                                   If there are any contests on the ballot that are not in the ballot style
                                   file entry, pretend in step~7(g) below
                                   that the CCVR for that (ballot, contest) pair
                                   showed a vote for the apparent winner of the contest.
                                   If there are any contests in the ballot style file entry that are not
                                   on the ballot, pretend in step~7(g) below
                                   that the ballot showed a vote for the apparent
                                   runner-up for that contest.
                          \item For each ballot $b_{\ell_j}$ in the sample,
                                   the election official reveals the value of $u_{\ell_j c}$
                                   for each contest $c$ on the ballot.
                          \item For each ballot in the sample, for each contest on that ballot,
                                   observers calculate $H(b_{\ell_j}, u_{\ell_jc})$ and find the entry in
                                   the CCVR file for contest $c$ that has that shrouded identifier.
                                   If the shrouded identifier is not in the CCVR file, pretend that
                                   the CCVR file showed that the voter had selected the apparent
                                   winner of contest $c$.
                                   Compare the voter's selection(s) according to the CCVR file to the voter's
                                   selection(s) according to a human reading of ballot $b_{\ell_j}$. 
                                   Find $e_{\ell_j}$, the largest number of votes by which any CCVR for ballot
                                   $b_{\ell_j}$ overstated the margin between any (winner, loser) 
                                   pair in any contest on ballot $b_{\ell_j}$.
                                   This number will be between $-2$ and $+2$.
                       \end{enumerate}
     \item[8.]  If no ballot in the sample has $e_{\ell_j} = 2$ and no more than $\lambda \mu n_0$
                    have $e_{\ell_j} = 1$, the audit stops.
                    (In this calculation, the value of $e_{\ell_j}$ should be counted as many times as 
                    $\ell_j$ occurs in the sample.)
     \item[9.]  Otherwise, calculate the Kaplan-Markov $P$-value, $P_{KM}$ according to 
                     equation~(9) in~\citet{stark09b,stark09d,stark10d}.\footnote{%
                 We consider only plurality voting here:  IRV is more complicated.
                 For each contest $c$, let $\cW_c$ be the indices of the apparent winners of the contest
                 and let $\cL_c$ be the indices of the apparent losers of the contest.
                 If $w \in \cW_c$ and $x \in \cL_c$, let $V_{wx}$ be the margin in votes between candidate 
                 $w$ and candidate $x$
                 according to the CCVR file for contest $c$.
                 For each candidate $k$ on ballot $\ell$, let $v_{\ell k}$ denote the number of votes for candidate $k$
                 on ballot $\ell$ according to the CCVR file and let $a_{\ell k}$ denote the number of votes on ballot $\ell$
                 for candidate $k$ according to a human reading of ballot $\ell$.
                 Let 
                 \begin{equation}
                      \epsilon_\ell \equiv \max_c \max_{w \in \cW_c, x \in \cL_c} (v_{\ell w}-a_{\ell w} - v_{\ell x} + a_{\ell x})/V_{wx}.
                 \end{equation}
                 Then 
                 \begin{equation}
                           P_{KM} \equiv \prod_{j=1}^n \frac{1 - 1/U}{1 - \frac{\epsilon_{\ell_j}}{2 \gamma/V}}.
                  \end{equation}
       }
                     If $P_{KM}$ is less than $\alpha$, the audit stops.
                     If $P_{KM}$ is greater than $\alpha$, the sample is expanded: 
                     Another random number $r_j$ is generated
                     and steps~7(c)--(g) are repeated. 
                     The value of $P_{KM}$ is updated to include the overstatement errors found in
                     the new draw.\footnote{%
                             Overstatements are calculated as step~7 above, including, in particular,
                             steps~7(e) and~7(g), which say how to treat failures to find ballots
                             or contests.
                     }
                     This continues until either $P_{KM} \le \alpha$ or
                     there have been $D$ draws.
                     In the latter case, all remaining ballots are counted by hand, revealing the true
                     outcome.
 \end{enumerate}
The next section establishes that this procedure in fact gives a risk-limiting audit.

\subsection{Proof of the risk-limiting property} \label{sec:proof}
If the ballot style file is correct and entries in the CCVR files are mapped
properly to voting opportunities on actual ballots, the only potential source of error
is that CCVR entries do not accurately reflect the voters' selections according to
a human reading of the ballot.
If that is the case, this is an ``ordinary'' risk-limiting audit, and the proof in~\citet{stark10d}
that the super-simple simultaneous method is risk-limiting applies directly.

Suppose therefore that the ballot style file or the mapping between ballots 
and CCVRs is faulty.
Recall that the super-simple simultaneous method assumes that no ballot can overstate any
margin by more than $2\gamma$ votes, where $\gamma > 1$.
There are seven cases to consider.
\begin{enumerate}
  \item  The ballot style file has more than one entry that corresponds to the same actual
            ballot, or more than one actual ballot corresponds to the same entry in the ballot
            style file.
            These faults are precluded by the uniqueness of the ballot identifiers and of the 
            recipes for locating the actual ballot with each identifier.
            
   \item More than one ballot identifier corresponds to the same shrouded entry (for different values
            of $u$).
            This is precluded by the binding property of $H$.

   \item The ballot style file contains identifiers that do not correspond to actual ballots,
            or claims that a ballot contains a contest that it does not actually contain.
            The biggest effect this could have on an apparent contest outcome is if the
            ballot that entry is supposed to match showed a vote for the runner-up
            in every missing contest, which is no greater than a two-vote change to any margin.
            Because the audit samples entries of the ballot style file with equal probability,
            this kind of error in an entry is just as likely to be revealed as any other.
            If such a ballot style file entry is selected for audit, steps~7(d) and 7(e) 
            treat it this worst-case way.
            
   \item The ballot style file claims that a ballot does not contain a contest 
            that it does contain.
            The biggest effect this could have on an apparent contest outcome is
            if the CCVR for that contest showed a vote for the apparent winner,
            which cannot change the margin by more than two votes, so the
            error-bound assumptions are satisfied.
            Because the audit samples entries of the ballot style file with equal probability,
            this kind of error in an entry is just as likely to be revealed as any other.
            If such a ballot style file entry is selected for audit, step~7(e) treats
            it this worst-case way.
            
   \item There are ballots whose identifiers do not appear in the ballot style file.
            Since there are the same number of ballots as entries in the ballot style file
            and the ballot identifiers in the ballot style file are unique,
            there must be ballot identifiers in the ballot style file that do not match
            any ballot.
            Hence, case~(3) holds.
            
   \item There are CCVRs for which the shrouded ballot identifier is not
            the identifier of any ballot.
            If the shrouded identifier matches an identifier in the ballot style file,
            we are in case~(3).
            Suppose therefore that the shrouded identifier does not match
            any in the ballot style file.
            Suppose this happens for contest $c$.
            The preliminary checks show that the ballot style file has 
            exactly $N_c$ entries for contest $c$ and that there
            are exactly $N_c$ entries in the CCVR file for contest $c$.
            Therefore, if there is such a CCVR, one of the ballot style file entries that lists contest $c$
            has an identifier that does not occur in shrouded form in the CCVR file for that contest.
            The largest effect this could have on contest $c$ is if the ``substituted'' CCVR entry
            reported a vote for the apparent winner; this cannot overstate the margin by more
            than two votes, so the audit's error-bound assumption still holds.
            Because the audit samples entries of the ballot style file with equal probability,
            this kind of error in a ballot style file entry is just as likely to be revealed as any other.
            If such a ballot style file entry is selected for audit, step~7(e) 
            treats it this worst-case way.
            
   \item The same ballot identifier appears in shrouded form more than once in a single
            CCVR file.
            As in the previous case, we know there are $N_c$ entries in the CCVR file for contest $c$
            and $N_c$ entries in the ballot style file that include contest $c$; moreover,
            the identifiers in the ballot style file are unique.
            Hence, there must be at least one entry in the ballot style file that lists contest $c$
            for which the ballot identifier does not appear in shrouded form in the CCVR file.
            We are therefore in case~(6).
\end{enumerate}

\section{Discussion} \label{sec:discussion}
Others have proposed election verification methods that involve a cryptographic
commitment by elections
officials to a mapping between ballots and 
CVRs~[E.K.~Rescorla, personal communication, 2011; 
R.L.~Rivest, personal communication, 2009; D.~Wallach, personal communication, 2010;
see also~\citet{adida06}].
However, we believe SOBA is the first method that requires only one commitment 
and that uses a risk-limiting audit to check whether the mapping is accurate 
enough to determine the correct winner.

We have said little about the requirement for a compliance audit.
In part, this is a definitional issue: 
Even if the audit trail is known to have been
compromised, it is our understanding that in many states, a full hand count of the
audit trail would still be the ``correct'' outcome, as a matter of law.
Hence, an audit to assess whether the audit trail was protected and preserved 
adequately for it to reflect the outcome according to how the voters 
cast their ballots is legally superfluous.
We consider this a shortcoming of current audit and recount laws.
Moreover, we doubt that any system can be $P$-resilient unless the election
and the data it generates satisfies particular conditions.
For instance, risk-limiting audits generally assume that the number of ballots
cast in all in each contest is known.
Such conditions should be checked.

We would advocate carrying out a compliance audit to assess whether the
procedures as followed in the election give reasonable assurance that the audit trail is 
trustworthy---sufficiently accurate to reflect the outcome according to how 
voters cast their ballots---and to assess whether any other preconditions of the 
risk-limiting audit hold.
The compliance audit should evaluate whether there is strong evidence
that the chain of custody of the ballots is intact, or whether it is plausible
that ballots were lost, ``found,'' altered, or substituted.
The compliance audit should confirm the values of $\{N_c\}$
by ballot accounting: confirming that the number of ballots printed
equals the number returned voted, unvoted, and spoiled, for each ballot type.

If the election passes the compliance audit, a risk-limiting audit can then
assess the accuracy of the reported result and would
have a large chance of correcting the apparent outcome if it is wrong 
(by examining the full audit trail). 
But if the election fails the compliance audit---that is, if we lack strong evidence
that the audit trail is reliable and that the preconditions for the risk-limiting 
audit are met---a $P$-resilient 
election framework should not declare any outcome at all.

For the method to be $P$-resilient, $H$ must be binding and we must know
$\{N_c\}$.
Because the election official discloses $H$ and the (fixed) length of the ballot identifiers,
we can determine whether $H$ is binding.
For the method to be privacy-preserving, $H$ must have the hiding property,
which will depend on how the salts are chosen
and how the CCVR files are organized.
If the salts can be discovered, inferred, or guessed, or if observers have
another way to reassemble whole-ballot CVRs from the CCVRs (for instance, if
the CCVRs are in the same ballot order across contests), 
voter privacy can be compromised.

\section{Conclusions}
SOBA makes possible a personally verifiable privacy-preserving $P$-resilient canvass framework.
It allows individuals to obtain strong firsthand\footnote{%
   For multi-jurisdictional contests, it might not be possible to conduct an audit in a single
   place and time.  
   If the audit step takes place in pieces in separate jurisdictions simultaneously, 
   firsthand knowledge might be impossible; one might need to trust 
   observers in other locations.
} 
evidence that apparent election outcomes 
either are correct in the first place, or are corrected by a 
risk-limiting audit before becoming
final, without unnecessary compromises to privacy.
After the procedure is complete, either all the outcomes are correct or an event with
probability less than $1-P$ has occurred.
The published data structures allow the public to check the consistency of
the apparent outcomes but
do not allow whole-ballot cast vote records to be reconstructed,
thereby preserving privacy.
When all the apparent contest outcomes are correct, 
gathering the evidence that the outcomes are right typically will require 
exposing only a small fraction of  ballots to observers, protecting privacy.
But the data structures and auditing protocol ensure that if the 
apparent outcome of one or more of the contests is wrong, there is a large chance 
of a full hand count of the audit trail to set the record straight.

\section{Acknowledgments}
This work was supported in part by NSF Grant CNS-05243 (ACCURATE).
We are grateful to Poorvi Vora for shepherding the paper and to anonymous referees for
helpful comments.
We are grateful to
Joseph Lorenzo Hall,
David Jefferson,
Neal McBurnett,
Dan Reardon,
Ronald L.~Rivest,
and
Emily Shen
for helpful conversations and comments on earlier drafts.

\bibliographystyle{apalike}
\bibliography{../../Bib/pbsBib}

\end{document}